\documentclass[preprint,showpacs,preprintnumbers,amsmath,amssymb]{revtex4}
\usepackage{graphicx}
\usepackage{slashed}
\usepackage{subfig}
\usepackage{float}
\usepackage[utf8]{inputenc}

\usepackage{color}
\definecolor{red}{rgb}{1,0,0}

\begin{document}

\title{Effective Majorana neutrino decay.}

\author{Luc\'{\i}a Duarte}
\email{lduarte@fing.edu.uy}
 \affiliation{Instituto de F\'{\i}sica, Facultad de Ingenier\'{\i}a,
 Universidad de la Rep\'ublica \\ Julio Herrera y Reissig 565,(11300) 
Montevideo, Uruguay.}

\author{Ismael Romero}
\author{Javier Peressutti}
\author{Oscar A. Sampayo}
\email{sampayo@mdp.edu.ar}

 \affiliation{Instituto de Investigaciones F\'{\i}sicas de Mar del Plata (IFIMAR)\\ CONICET, UNMDP\\ Departamento de F\'{\i}sica,
Universidad Nacional de Mar del Plata \\
Funes 3350, (7600) Mar del Plata, Argentina}

\begin{abstract}
We study the decay of heavy sterile Majorana neutrinos according to the interactions obtained from an effective general theory. We describe the two and three-body decays for a wide range of neutrino masses. The results obtained and presented in this work could be useful for the study of the production and detection of these particles in a variety of high energy physics experiments and astrophysical observations. We show in different figures the dominant branching ratios and the total decay width.
 
\end{abstract}

\maketitle

\section{\bf Introduction}

The discovery of neutrino oscillations has been one of the most spectacular new results in high energy physics, and so far is the only compelling experimental evidence of the existence of physics beyond the Standard Model. The sub-$e$V left handed neutrino masses required by neutrino oscillation data are very difficult to generate just by the addition of right handed neutrinos to the Standard Model, as the Yukawa couplings should be very small compared to those of the other particles. The introduction of intermediate fermion heavy particles which are singlets under the $SM$ gauge group -the right handed Majorana neutrinos- allows for the generation of light neutrino masses by the seesaw mechanism \cite{Minkowski:1977sc, Mohapatra:1979ia, Yanagida:1980xy, GellMann:1980vs, Schechter:1980gr, Kayser:1989iu}.

For the conventional seesaw scenarios often studied, the light neutrino masses are inversely proportional to an unknown lepton number violating large scale $M_N$ such that  $m_{\nu} \sim m_D^2/M_N$ where $m_D$ is a Dirac mass connected with the Yukawa coupling by $m_D=Y v/\sqrt{2}$, being $v$ the Higgs field vacuum expectation value. For Yukawa couplings of order $Y \sim 1$ we need a Majorana mass scale of order $M_{N} \sim 10^{15} GeV$ to account for a light $\nu$ mass compatible with the current neutrino data ($m_{\nu}\sim 0.01 e$V). This scenario clearly leads to the decoupling of the heavy Majorana neutrino $N$. However, for smaller Yukawa couplings of the order $Y \sim 10^{-8} - 10^{-6}$, sterile neutrinos with masses $M_N \sim (1-1000) GeV$ could exist. Any way, in the simplest Type-I seesaw scenario with sterile neutrinos, this leads to a too small left-right neutrino mixing \cite{deGouvea:2015euy, Deppisch:2015qwa, delAguila:2008ir}, $U_{lN}^2 \sim m_{\nu}/M_N \sim 10^{-14}-10^{-10}$. These values are several orders of magnitude smaller than the neutrinoless double beta decay ($0\nu\beta\beta$) or collider bounds, as will be shown later. 

Thus, as it was explained in \cite{delAguila:2008ir}, the detection of Majorana neutrinos would be a signal of physics beyond the minimal seesaw mechanism leading to the well known $\nu SM$ lagrangian, and its interactions could be better described in a model independent approach based on an effective theory, considering a scenario with only one Majorana neutrino $N$ and negligible mixing with the $\nu_{L}$.      

The possibilities of discovering heavy Majorana neutrinos have been and are still extensively investigated, for example involving production and decay in $e^+e^-$ and $e^- P$ colliders \cite{Ma:1989jpa, Buchmuller:1991tu, Hofer:1996cs, Peressutti:2011kx, Blaksley:2011ey, Duarte:2014zea, Antusch:2015mia, Banerjee:2015gca}, in $e^- \gamma$ and $\gamma \gamma$ colliders \cite{Bray:2005wv, Peressutti:2001ms, Peressutti:2002nf}, in hadron colliders via lepton number violating dilepton signals \cite{Keung:1983uu, Datta:1993nm, Almeida:2000pz, Atre:2009rg, delAguila:2007qnc, delAguila:2008ir, Kovalenko:2009td, Alva:2014gxa, Das:2015toa, Dib:2015oka, Izaguirre:2015pga}, and recently including new production mechanisms \cite{Das:2016hof, Degrande:2016aje, Dev:2013wba}. Some searches are currently being performed in the LHC \cite{Khachatryan:2015gha, Aad:2015xaa, Aad:2011vj, ATLAS:2012ak}. Also, we can mention recent inverse seesaw mechanism (ISS) \cite{Mohapatra:1986aw, Mohapatra:1986bd, Bernabeu:1987gr} heavy neutrino production studies in collider contexts \cite{Arganda:2015ija, Das:2012ze, Das:2014jxa, Gluza:2016qqv}.

The study of sterile Majorana neutrino decays is an issue of great interest in different areas of high energy physics. Besides the mentioned detection in colliders by lepton number violation, other kinds of searches exploiting the displaced vertex and delayed photons techniques have been proposed and are taking place at the LHC \cite{Aad:2013oua, Helo:2013esa, Aad:2015rba, Aad:2014tda, Gago:2015vma, Biswas:2010yp, Antusch:2016vyf, Batell:2016zod}. Also, searches in neutrino telescopes like Ice Cube \cite {Pagliaroli:2015rca,Masip:2011qb} have been proposed, and the new decay modes and their relation with the explanation of several anomalies as the sub-horizontal events detected by SHALON or the anomaly in MiniBoone \cite{Ross-Lonergan:2015pna,Masip:2012ke} are being investigated \cite{Dib:2011hc, Duarte:2015iba}. In astrophysical environments, the cosmic and the diffuse supernova neutrino backgrounds can be used to probe possible radiative decays and other decay modes of cosmological interest \cite{Fogli:2005fxu, Kim:2011ye}. 

With these motivations in mind, in this work we study the decays of heavy Majorana neutrinos in a general, model-independent approach in the context of an effective theory. In section \ref{sec:L_ope} we present the effective operators and the analytical decay widths obtained for the different two-body and three-body channels. In section \ref{sec:Num} we present our numerical results for the found decay modes, and discuss the bounds imposed on the values for the effective couplings. Our final remarks are made in section \ref{sec:fin}. The complete effective lagrangian and fermionic decay modes are left for the appendix.

\section{Effective operators and Decay widths}\label{sec:L_ope}
  
In this paper we consider the decays of a right handed Majorana neutrino $N$. As it is a $SM$ singlet, the only possible renormalizable interactions with the $SM$ fields could occur via the the Yukawa coupling, which as we mentioned earlier, must be very small if the $\nu SM$ is to reproduce the observed tiny $\nu_{L}$ masses. In an alternative approach, in this paper we consider that the sterile $N$ interacts with the standard light neutrinos by effective operators of higher dimension. We consider this effective interaction to be dominant compared to the mixing via the Yukawa couplings, so we depart from the traditional viewpoint in which the sterile neutrinos mixing with the standard neutrinos is assumed to govern the production and decay mechanisms for the $N$.

In this approach we parameterize the effects of new physics beyond the
Standard Model by a  set of effective operators $\mathcal{O}$
constructed with the SM and the Majorana neutrino fields
and satisfying the Standard Model $SU(2)_L \otimes U(1)_Y$ gauge
symmetry \cite{Wudka:1999ax}.  The effect of these operators is suppressed by inverse
powers of the new physics scale $\Lambda$ -which is not necessarily related to the mass $m_{N}$- for which we take the value $\Lambda = 1 \; TeV$ \cite{Wudka:2009zz}.

The total lagrangian is organized as
follows:

\begin{eqnarray}
\mathcal{L}=\mathcal{L}_{SM}+\sum_{n=6}^{\infty}\frac1{\Lambda^{n-4}}\sum_i
\alpha_i \mathcal{O}_i^{(n)}
\end{eqnarray}

For the considered operators we follow \cite{delAguila:2008ir} starting
with a rather general effective lagrangian density for the
interaction of right handed Majorana neutrinos $N$ with leptons and
quarks. All the operators we list here
are of dimension $6$ and could be generated at tree-level in the
unknown fundamental high energy theory. The first subset includes operators with scalar and vector bosons (SVB), 

\begin{eqnarray} \label{eq:ope1}
\mathcal{O}_{LN\phi}=(\phi^{\dag}\phi)(\bar L_i N \tilde{\phi}),
\;\; \mathcal{O}_{NN\phi}=i(\phi^{\dag}D_{\mu}\phi)(\bar N
\gamma^{\mu} N), \;\; \mathcal{O}_{Ne\phi}=i(\phi^T \epsilon D_{\mu}
\phi)(\bar N \gamma^{\mu} e_i)
\end{eqnarray}

and a second subset includes the baryon-number conserving four-fermion contact terms:

\begin{eqnarray} \label{eq:ope2}
\mathcal{O}_{duNe}=(\bar d_i \gamma^{\mu} u_i)(\bar N \gamma_{\mu}
e_i) &,& \;\; \mathcal{O}_{fNN}=(\bar f_i \gamma^{\mu} f_i)(\bar N
\gamma_{\mu}
N), \\
\mathcal{O}_{LNLe}=(\bar L_i N)\epsilon (\bar L_i e_i)&,& \;\; \nonumber
\mathcal{O}_{LNQd}=(\bar L_i N) \epsilon (\bar Q_i d_i), \cr
\mathcal{O}_{QuNL}=(\bar Q_i u_i)(\bar N L_i)&,& \;\;
\mathcal{O}_{QNLd}=(\bar Q_i N)\epsilon (\bar L_i d_i),\cr
\mathcal{O}_{LN}=|\bar N L_i|^2&&
\end{eqnarray}

where $e_i$, $u_i$, $d_i$ and $L_i$, $Q_i$ denote, for the family
labeled $i$, the right handed $SU(2)$ singlet and the left-handed
$SU(2)$ doublets, respectively.

In addition, there are operators generated at one-loop level in the underlying full theory whose coefficients are naturally suppressed by a factor $1/16\pi^2$\cite{delAguila:2008ir, Arzt:1994gp}:
\begin{eqnarray} 
\mathcal{O}^{(5)}_{NNB} & = & \bar N \sigma^{\mu\nu} N^c B_{\mu\nu}, \cr
\mathcal{O}_{ N B} = (\bar L \sigma^{\mu\nu} N) \tilde \phi B_{\mu\nu} , &&
\mathcal{O}_{ N W } = (\bar L \sigma^{\mu\nu} \tau^I N) \tilde \phi W_{\mu\nu}^I , \cr
\mathcal{O}_{ D N} = (\bar L D_\mu N) D^\mu \tilde \phi, &&
\mathcal{O}_{ \bar D N} = (D_\mu \bar L N) D^\mu \tilde \phi \ .
\label{eq:ope3}
\end{eqnarray}
\subsection{Two-body decays}

\begin{figure*}[h!]
\centering
\includegraphics[width=\textwidth]{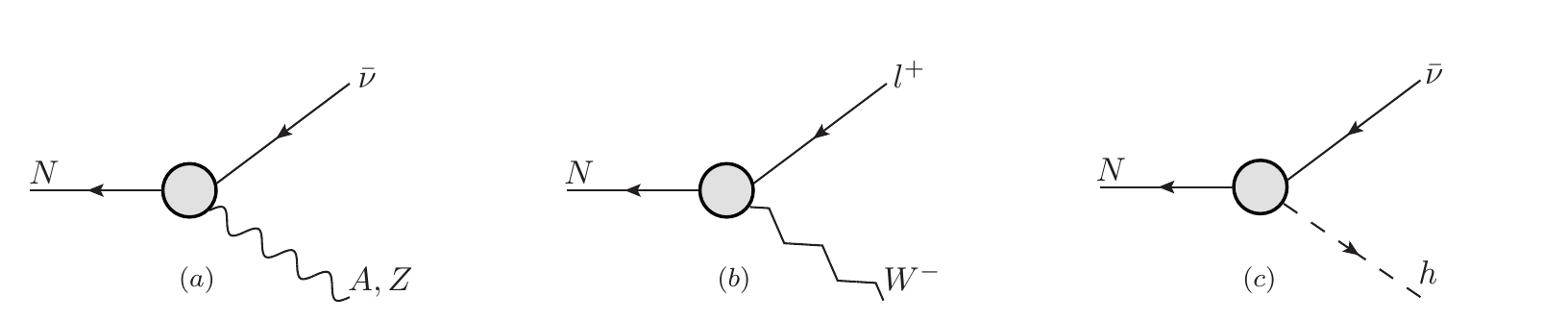}
\caption{\label{fig:N2bdec}  Two-body decays with gauge bosons and Higgs field.}
\end{figure*}

The two-body decay channels for the heavy Majorana neutrino $N$ are shown in Fig.(\ref{fig:N2bdec}). They receive contributions from the lagrangian terms originated in operators involving gauge bosons and the Higgs field, presented in \eqref{eq:ope1} and \eqref{eq:ope3}, that lead to the effective lagrangian presented in \ref{leff_svb} and \ref{leff_1loop}.

The analytical expressions obtained for the decay widths of channels $N \rightarrow \nu Z$, $N\rightarrow l^{+} W^{-}$, $N\rightarrow \nu h$  shown in Fig.(\ref{fig:N2bdec}) are:

\begin{eqnarray*}
\Gamma^{N\rightarrow \nu_i Z}&&=\left(\frac{m_N}{128\pi}\right)\left(\frac{m_N}{\Lambda}\right)^4(1-y_Z)^2\left[(\alpha_{L_4}^{(i)}-\alpha_{L_2}^{(i)})^2(1-y_Z)^2+ \right. \nonumber \\
&& 8(\alpha_{L_4}^{(i)}-\alpha_{L_2}^{(i)})(\alpha_{L_3}^{(i)}c_W-\alpha_{L_1}^{(i)}s_W)(1-y_Z)\sqrt{y_Z y_v}+\nonumber \\ 
&& \left. 16(2+y_Z)y_V(\alpha_{L_3}^{(i)}c_W-\alpha_{L_1}^{(i)}s_W)^2 \right]
\end{eqnarray*}
with $y_Z=m_Z^2/m_N^2$ and $y_v=v^2/m_N^2$.
\begin{eqnarray*}
\Gamma^{N\rightarrow l_i W}=\frac{m_N}{32\pi}\left( \frac{m_N}{\Lambda}\right)^4 \alpha_{W}^{(i)}(1-y_W)^2(1+2y_W)y_v
\end{eqnarray*}
with $y_W=m_W^2/m_N^2$.

\begin{eqnarray*}
\Gamma^{(N \rightarrow \nu_i h)}&=&\frac{9 m_N}{128 \pi} \left(\frac{v}{\Lambda}\right)^4
\alpha_{\phi}^{(i)2} (1-y_{h}) 
\end{eqnarray*}
with $y_{h}={m_{h}^2}/{m_N^2}$. 

Finally we have the decay mode to a photon and an ordinary neutrino $N\rightarrow \nu A$:
\begin{eqnarray} \label{photon_neutrino}
\Gamma^{N\rightarrow \nu_i A}=\frac1{4\pi}\left(\frac{v^2}{m_N}\right)\left(\frac{m_N}{\Lambda}\right)^4(\alpha_{L_1}^{(i)}c_W+\alpha_{L_3}^{(i)}s_W)^2
\end{eqnarray}
This decay mode leads to an interesting phenomenology, part of which was disused in \cite{Duarte:2015iba}.

It is important to take into account here that the $W$ and $h$ resonant contributions to other decays, as can be seen in Figs. \ref{fig:N3bdec_3f} (a) and (b) were already included in those decays and will not be added to the total width.

\subsection{Three-body decays}

The three-body decays of the heavy Majorana neutrino $N$ involving gauge bosons and the Higgs field receive contributions from the lagrangians presented in \eqref{leff_svb} and \eqref{leff_1loop}, whereas the decays to three fermions come also from the operators presented in \eqref{eq:ope2}. The effective model we are working with also gives tree-level contributions to four-body decays, but as their contributions are very small, they are not presented in this work.

\begin{figure*}[h!]
\centering
\includegraphics[width=\textwidth]{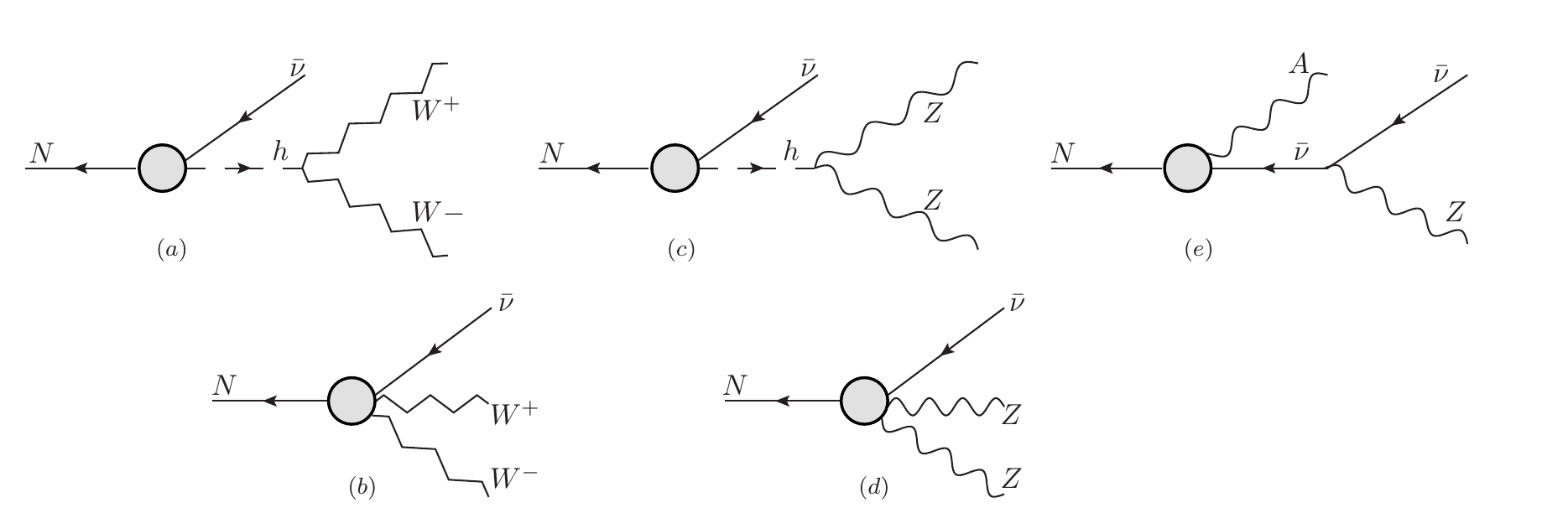}
\caption{\label{fig:N3bdec_gb_n}  Three-body decays with two gauge bosons and ordinary neutrinos.}
\end{figure*}

The three-body decay channels involving gauge bosons and ordinary neutrinos are shown in Fig.(\ref{fig:N3bdec_gb_n}); (a) and (b) $N\rightarrow \nu W^+ W^-$, (c) and (d) $N\rightarrow \nu Z Z$, and (e) $N \rightarrow \nu Z A$. 

The analytical expressions for the decay widths are:
\begin{eqnarray*}\label{WWnu}
\frac{d\Gamma^{N \rightarrow \nu_i W^+ W^-}}{dx}=&&\frac{m_N}{6144 \pi^3}\left(\frac{m_N}{\Lambda}\right)^4 \frac{x^2}{(1-x)^2 y_W}
((1-x)(1-x-4y_W))^{1/2}\times
\nonumber\\
&&\left[16\alpha_{L_3}^{(i)2} (3-x)((1-x)^2+4(1-x)y_W-8y_W^2) +\right.
\nonumber\\
&&\left. 3 \mid\tilde{\alpha}^{(i)}\mid^2 g^2 (1-x)((1-x)^2-4(1-x)y_W+12 y_W^2)\right]
\end{eqnarray*}
where $\tilde{\alpha}^{(i)}=\alpha_{L_4}^{(i)}+\frac32 \alpha_{\phi}^{(i)} \frac{y_v}{1-x-y_{h}+i \sqrt{y_{h} y_{\Gamma_W}}},~ y_{\Gamma_W}={\Gamma_W^2}/{m_N^2}$.
In this process we discard the $N\rightarrow l W$ followed by the $l \rightarrow \nu W$ $SM$ vertex contribution, because the amplitude is proportional to the intermediate lepton mass, and thus negligible in comparison with the diagrams shown in Fig.(\ref{fig:N3bdec_gb_n}) (a) and (b).

\begin{eqnarray*}\label{ZZnu}
\frac{d\Gamma^{N \rightarrow \nu_i Z Z}}{dx}&&=\frac{m_N}{64 \pi^2}\left(\frac{m_N}{\Lambda}\right)^4 
\frac{\alpha^{emg}}{s_{2W}^2} y_Z \left[\left(\alpha_{L_4}^{(i)}+\frac{3\alpha_{\phi}^{(i)}y_v(1-x-y_{h})}
{(1-x-y_{h})^2+y_{h}y_{\Gamma_{h}}}\right)^2 \right. 
\nonumber\\ 
&&\left. + \left(\frac{3\alpha_{\phi}^{(i)}y_v \sqrt{y_{h}y_v}}{(1-x-y_{h})^2+y_{h}y_{\Gamma_{h}}}\right)^2 \right]
\times 
\nonumber\\
&&\frac{x^2}{(1-x)}\left(2+\frac{(1-x-2y_Z)^2}{4 y_Z^2}\right)
((1-x-2 y_Z)^2-4 y_Z^2)^{1/2}
\end{eqnarray*}
with $\alpha^{emg}$ being the electromagnetic constant, and $y_{\Gamma_{h}}={\Gamma_{h}^2}/{m_N^2}$.
\begin{eqnarray*}\label{ZAnu}
\frac{d\Gamma^{N \rightarrow \nu_i Z A}}{dx}=\frac{m_N}{32\pi^3}\left(\frac{m_N}{\Lambda}\right)^4 \left(c_W \alpha_{L_1}^{(i)}
+s_W \alpha_{L_3}^{(i)}\right)^2 \frac{(1-x+2 y_Z)(1-x-y_Z)x^3}{(1-x)^3}
\end{eqnarray*}

%
\begin{figure*}[h!]
\centering
\includegraphics[width=0.7\textwidth]{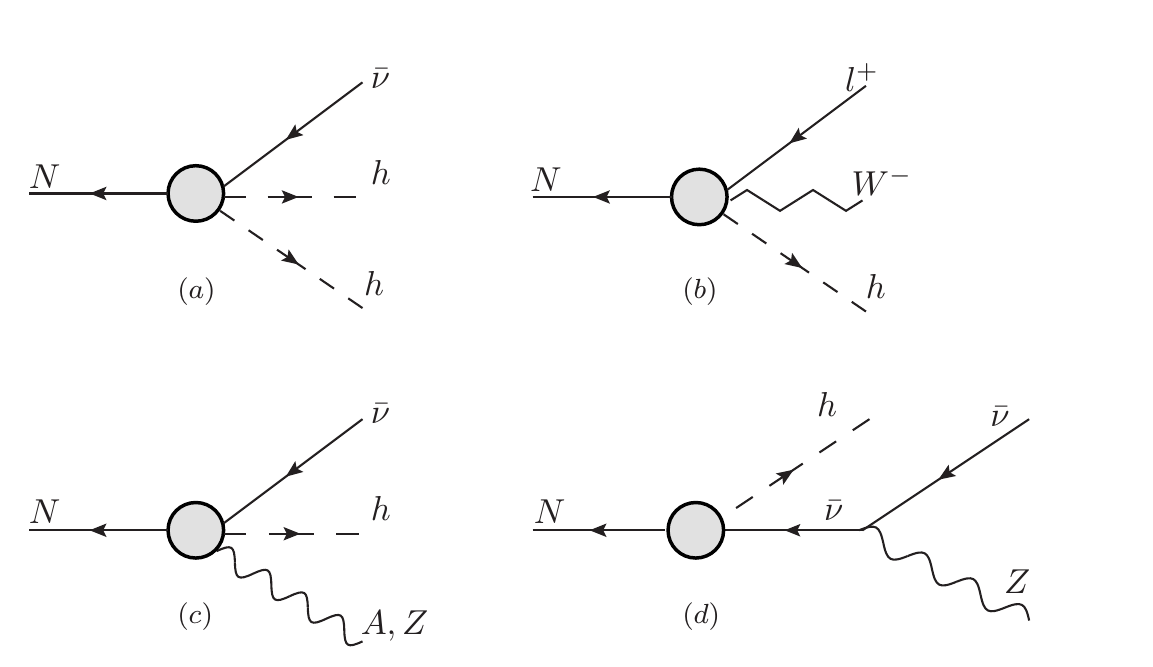}
\caption{\label{fig:N3bdec_h_gb}  Three-body decays with gauge bosons and Higgs field.}
\end{figure*}

The three-body channels with Higgs fields in the final state are shown in Fig.(\ref{fig:N3bdec_h_gb}) (a) $N \rightarrow \nu h h$, (b) $N\rightarrow l^{+} W^{-} h$, (c) $N\rightarrow \nu h A$ and (d) $N\rightarrow \nu h Z$. The obtained expressions for the decay widths are:
\begin{eqnarray*}
\frac{d\Gamma^{N \rightarrow \nu_i h h}}{dx}=\frac{18 \alpha_{W}^{(i)2}m_N}{2048 \pi^3}\left(\frac{v}{m_N}\right)^2
\left(\frac{m_N}{\Lambda}\right)^4 x^2 \frac{((1-x-2y_h)^2-4 y_h^2)^{1/2}}{(1-x)}
\end{eqnarray*}
with $0\leq x \leq 1-4 y_h$.

\begin{eqnarray*}
&&\frac{d\Gamma^{N \rightarrow l_i W h}}{dx}=\frac{m_N \alpha_W^{(i)2}}{768 \pi^3}\left(\frac{m_N}{\Lambda}\right)^4
((1-x-y_h)^2-2(1-x+y_h)y_W+4 y_W^2)^{1/2} \times
\nonumber\\
&&\left[(3-x)\left((1-x-y_h)^2+y_W^2\right)+(6-3 y_h+x(-11+5x+y_h))y_W \right] \frac{x^2}{(1-x)^3}
\end{eqnarray*}

\begin{eqnarray*}
 \frac{d\Gamma^{N \rightarrow \nu_i A h}}{dx}=\frac{m_N}{192 \pi^3}\left(\frac{m_N}{\Lambda}\right)^4 \left(c_W \alpha_{L_1}^{(i)}
+s_W \alpha_{L_3}^{(i)}\right)^2 \frac{(1-x-y_h)^3 x^3}{(1-x)^3} 
\end{eqnarray*}
with $0\leq x\leq1-y_h$.

\begin{eqnarray*}
 \frac{d\Gamma^{N \rightarrow \nu_i Z h}}{dx}&&=\frac{9 ~m_N ~g^2}{2^{14} \pi^3}
 \left(\frac{v}{\Lambda}\right)^4  \alpha_{\phi}^{(i)2}  \times
 \nonumber\\
&&\left[ \frac{(2-x)(1-x+y_h+2y_z)(1-x+y_h-y_z)(x^2-4y_h)^{1/2}}{y_z(1-x+y_h)}\right]
\end{eqnarray*}
with $2 \sqrt{y_h} \leq x \leq 1+y_h-y_z$.
This decay width is obtained from the diagram shown in Fig.(\ref{fig:N3bdec_h_gb})(d), as this contribution involves a tree-level vertex coming from the lagrangian \eqref{leff_svb} and a $SM$ vertex, and is dominant comparing to the one-loop level term coming from the lagrangian \eqref{leff_1loop}, that would give a vertex as the one shown in Fig.(\ref{fig:N3bdec_h_gb}) (c). 

\begin{figure*}[h!]
\centering
\includegraphics[width=\textwidth]{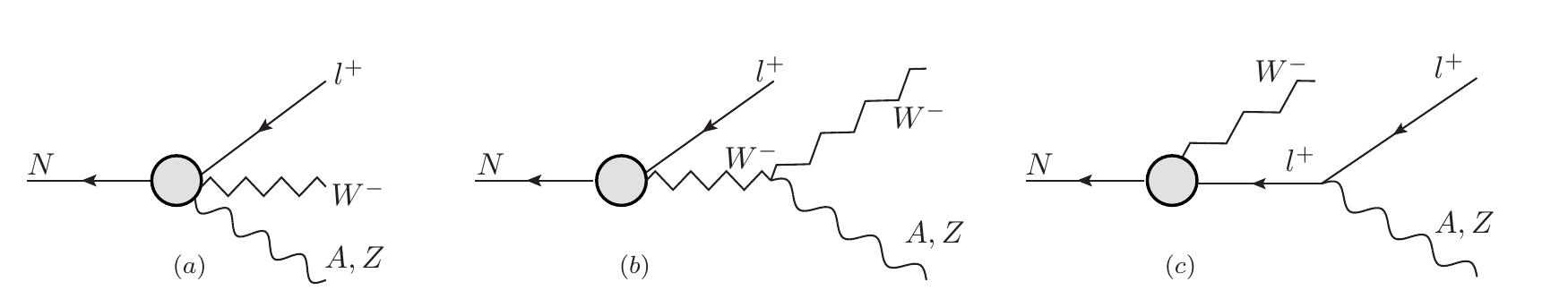}
\caption{\label{fig:N3bdec_gb_l} Three-body decays with two gauge bosons and charged leptons.}
\end{figure*}

The three-body decay channels with two gauge bosons and charged leptons in the final state are shown in Fig.(\ref{fig:N3bdec_gb_l}) $N\rightarrow l^{+} W^{-} A,Z$. We cannot obtain analytical expressions for these decay widths, and we have done numerical integrations of the phase space in the usual way using the numerical routine RAMBO \cite{Kleiss:1985gy}.

Some of the three-body decays involving only fermions in the final state come from the four-fermion contact operators presented in \eqref{eq:ope2}. These operators lead to the tree-level lagrangian in \eqref{leff_4-f}.

\begin{figure*}[h!]
\centering
\includegraphics[width=\textwidth]{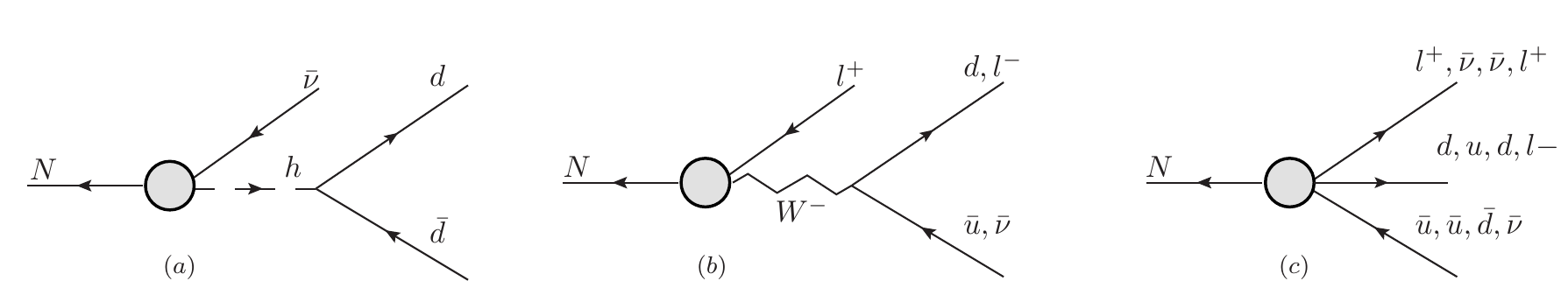}
\caption{\label{fig:N3bdec_3f}  Majorana neutrino decaying to three fermions. }
\end{figure*}

The partial decay widths of a heavy Majorana neutrino $N$ decaying to three fermions were calculated including the contributions in the effective lagrangians \eqref{leff_svb} and \eqref{leff_4-f}. The decay channels are shown in Fig.(\ref{fig:N3bdec_3f}). 
As was previously mentioned, the diagrams (a) and (b) show the resonant contributions coming from two-body decays to $W$ and $h$ bosons. The analytical expressions obtained were presented in our previous work \cite{Duarte:2015iba}, and for completeness we display them in the appendix \ref{Fermionic}. 

\section{Numerical branching ratios and total decay width}\label{sec:Num}

\begin{figure*}[!h]
\centering
\subfloat[Three fermions and neutrino-photon decay channels.]{\label{fig:3f_nug}\includegraphics[width=0.72\textwidth]{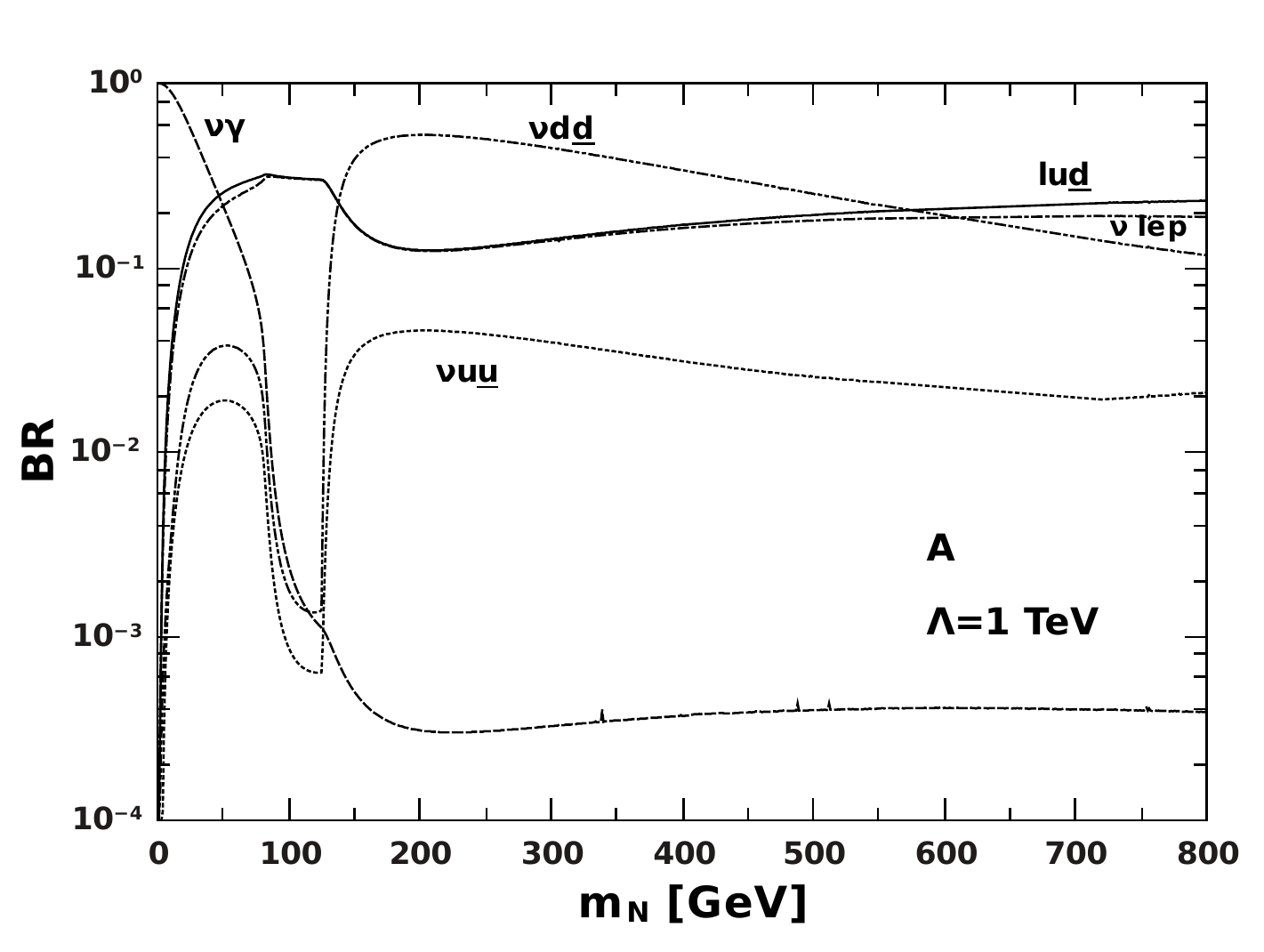}}

\subfloat[Massive gauge and Higgs boson decay channels]{\label{fig:G_h_b}\includegraphics[width=0.72\textwidth]{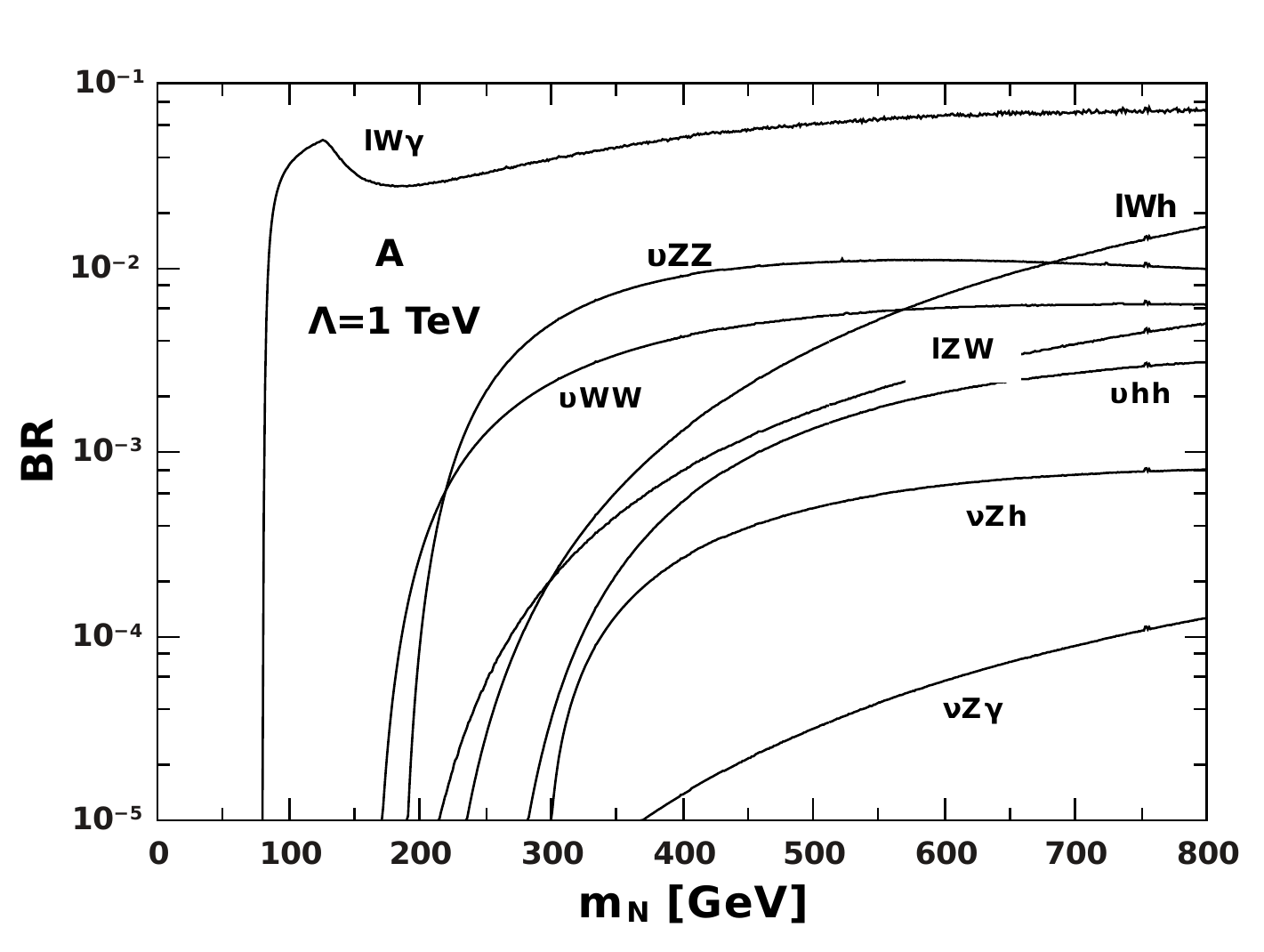}}
 \caption{\label{fig:br_majo0}The Branching ratios for the Majorana neutrino decay in the set {\bf A} considering the 
 sum of families.  }
\end{figure*}

\begin{figure*}[h!]
\centering
\includegraphics[width=0.7\textwidth]{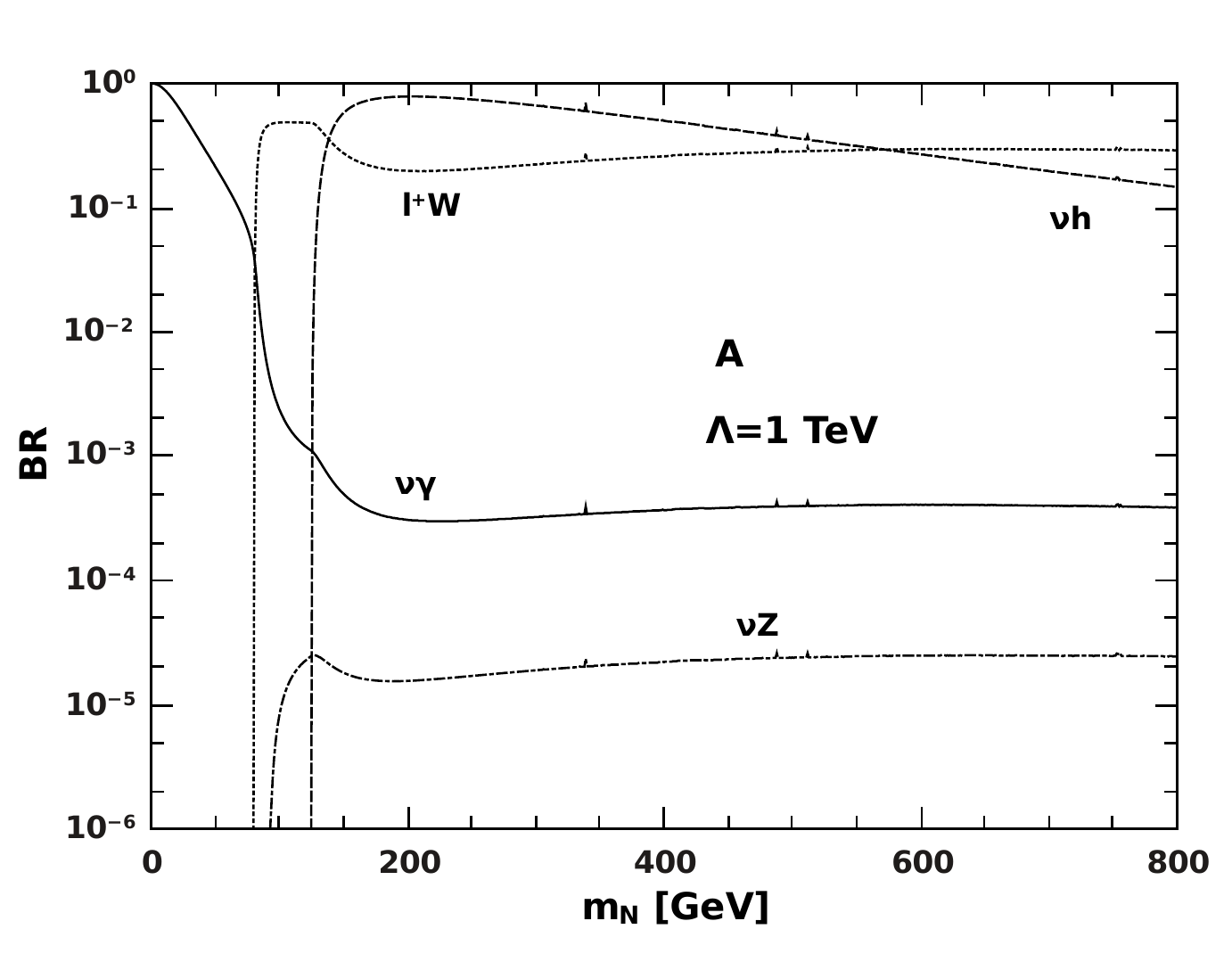}
 \caption{\label{fig:br_majo1}The branching ratios for two-body decays considering the sum over families.}
\end{figure*}

The numerical results for the Majorana neutrino branching ratios and total decay width are presented in the following. In Figs.(\ref{fig:br_majo0}) and (\ref{fig:br_majo1}) we show the results for the branching ratios for the different decay channels found in the previous section. We display the branching ratios as a function of the Majorana neutrino mass $m_N$, calculated for different numerical values of the constants $\alpha^i_{\mathcal O}$. In all the following results, when ordinary neutrinos are present in the final states, we sum the contributions of the neutrino and antineutrino channels. It is important to realize that, as we explained in the introduction, we are neglecting the contributions of the $N-\nu_{L}$ mixings compared to the effective interactions. In this condition the effective contributions to the branching ratios that we present here are dominant for the scale $\Lambda=1$ T$e$V considered and for the values of $\alpha$ allowed by experimental data, as will be explained in the next section.

The values for the coupling constants $\alpha$ are limited by bounds coming from electroweak precision data ($EWPD$) and $0\nu\beta\beta$-decay data. In order to simplify the discussion we consider just two numerical sets of values: in set {\bf A} the couplings associated to the operators that contribute to the $0\nu\beta\beta$-decay are restricted to the corresponding bound, $\alpha^{bound}_{0\nu\beta\beta}$ \eqref{eq:0nubb} and the other constants are restricted to the bound determined by $EWPD$ $\alpha^{bound}_{EWPD}$ \eqref{eq:EWPD} (see next section). In the case of the set {\bf B} all the couplings are restricted to the $0\nu\beta\beta$ bound $\alpha^{bound}_{0\nu\beta\beta}$ which is the most stringent. The branching ratios, being quotients between partial widths, take very similar numerical values for both sets.

In Fig.(\ref{fig:3f_nug}) we present the branching ratios of the three-fermion and photon-neutrino channels for the couplings set {\bf A}. These are the only open channels for Majorana neutrino masses below $m_W$. As it can be seen, for low $m_N$ the dominant mode is the decay of $N$ to a photon and a neutrino. 
Taking the values of the couplings $\alpha^{(i)}$ to be equal for every family $i$, and also for every tree-level coupling $\alpha^{tree}$, and taking the one-loop generated couplings as $\alpha^{1-loop}=\alpha^{tree}/16\pi^2$, we derived an approximated expression for the ratio between the widths $\Gamma (N \rightarrow \nu(\bar \nu) A)$ in \eqref{photon_neutrino} and $\Gamma (N\rightarrow l^{+} \bar u d)$ in \eqref{dquarks_1}:
\begin{eqnarray*}
 \frac{\Gamma^{(N \rightarrow \nu(\bar \nu) A)}}{\Gamma^{(N\rightarrow l^{+} \bar u d)}}\rightarrow \frac{2}{15 \pi} \left(\frac{v}{m_{N}}\right)^2 \left( c_{W}+s_{W}\right)^2
\end{eqnarray*}
This limiting value explains the behavior found in Fig.(\ref{fig:3f_nug}), showing the neutrino plus photon decay channel is clearly dominating for low $m_N$ (in \cite{Duarte:2015iba} we verify that this channel still dominates over the decay of $N$ to QCD-mesons like pions). This is an interesting fact since we have a new source of photons in astrophysical environments. The implications of this new channel for the MiniBoone \cite{AguilarArevalo:2008rc} and SHALON \cite{Sinitsyna:2013hmn} anomalies were discussed in our previous work \cite{Duarte:2015iba}.

Fig.(\ref{fig:G_h_b}) shows the massive gauge and Higgs boson decay channels for the $N$. The branching ratio for the $N\rightarrow \nu h A$ mode is smaller than $1\times10^{-7}$ and is not visible in the plot. 

For completeness we present the branching ratios for the two-body decay channels. As we explained in the previous section, the $N\rightarrow \nu h, W$ channels are not included in the total width, as their contribution has already been taken into account in the channels where the $W$ and $h$ bosons participate as intermediate resonant states.

\begin{figure*}[h!]
\begin{center}
\includegraphics[width=0.7\textwidth]{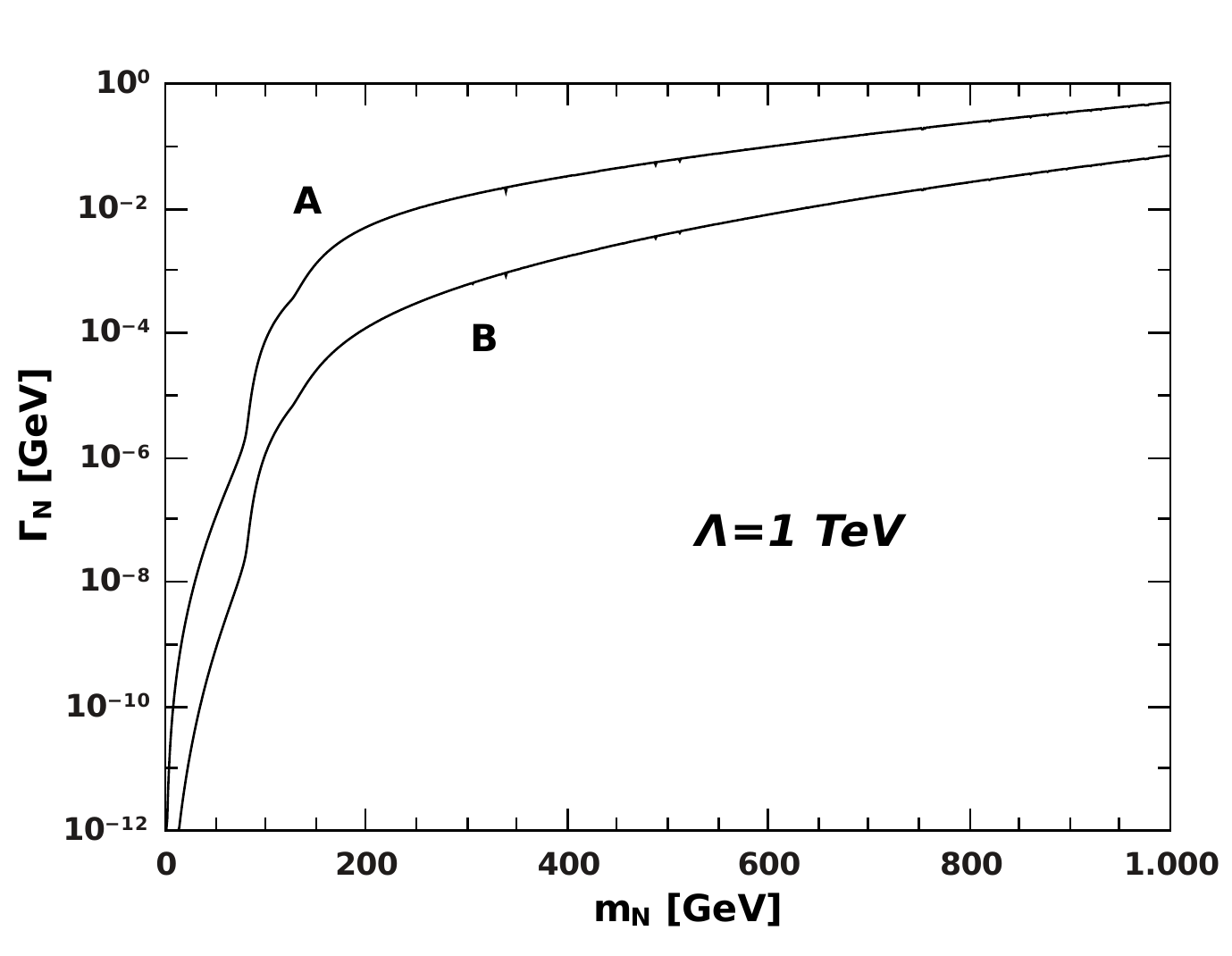}
\caption{\label{fig:width} Total decay width with coupling constants  in the set {\bf A} (solid line) and  set {\bf B} (dashed line), and $\Lambda = 1\,$T$e$V.}
\end{center}
\end{figure*}

Finally, Fig.(\ref{fig:width}) shows the total decay width dependence on the mass $m_N$ for both coupling sets considered, and again a sum over families and channels with particles and antiparticles in the final state is performed.  


\subsection{Bounds on the couplings $\alpha^i_{\mathcal{O}}$}\label{Bounds}

The effective couplings $\alpha^i_{\mathcal{O}}$ can be bounded exploiting the existing constraints coming from neutrinoless double beta decay ($0\nu\beta\beta$), electroweak precision data tests ($EWPD$), low energy observables as rare lepton number violating (LNV) meson decays and direct collider searches, including $Z$ decays.  
We explain here in detail how we take account of the existing bounds for the sterile-active neutrino mixings and turn them into constraints on the effective couplings $\alpha^i_{\mathcal{O}}$. In the literature, the bounds are generally imposed on the parameters representing the mixing between the sterile and active left-handed ordinary neutrinos. Very recent reviews \cite{deGouvea:2015euy, Deppisch:2015qwa, Antusch:2015mia, Drewes:2015iva} summarize in general phenomenological approaches the existing experimental bounds, considering low scale minimal seesaw models, parameterized by a single heavy neutrino mass scale $M_{N}$ and a light-heavy mixing $U_{lN}$, with $l$ indicating the lepton flavor.
In the effective lagrangian framework we are studying, the heavy Majorana neutrino couples to the three fermion family flavors with couplings dependent on the new ultraviolet physics scale $\Lambda$ and the constants $\alpha^{(i)}_{\mathcal O}$. We can interpret the current bounds comparing our couplings with the general structure usually taken for the interaction between heavy neutrinos with the standard gauge bosons \cite{delAguila:2006dx, delAguila:2008iz}: 
\begin{eqnarray}
\label{lw_lz}
 \mathcal L_W = -\frac{g}{\sqrt{2}}  \overline l \gamma^{\mu} U_{lN} P_L N W_{\mu} + h.c. \;\ 
 \mathcal L_Z = -\frac{g}{2 c_{W}} \overline \nu_{L} \gamma^{\mu} U_{lN} P_L N Z_{\mu} + h.c.
\end{eqnarray}
The operators presented in \eqref{eq:ope1} lead to a term in the effective lagrangian \eqref{leff_svb} that can be compared to the interaction in \eqref{lw_lz} for the weak charged current, giving a relation between the coupling $\alpha^{(i)}_{W}$ and the mixing $U_{lN}$: $ U_{l_{i}N}\simeq \frac{\alpha^{(i)}_{W}v^2}{2\Lambda^2}$ \cite{delAguila:2008ir}. Nevertheless, as we are neglecting the $N-\nu_{L}$ neutrino mixing, no operators lead to a term that can be directly related -with the same Lorentz-Dirac structure- to the interaction in \eqref{lw_lz} for the neutral current (nor at tree or one-loop level). Some terms in the lagrangian \eqref{leff_1loop} contribute to the $ZN\nu$ coupling, but as they are generated at one-loop level in the ultraviolet underlying theory, they are suppressed by a $1/16 \pi^2$ factor.
In consequence, we take a conservative approach and in order to keep the analysis as simple as possible -but with the aim to put reliable bounds on our effective couplings- in this work we relate the mixing angle between light and heavy neutrinos ($U_{eN}$, $U_{\mu N}$, $U_{\tau N}$) with the couplings as $U \simeq \frac{\alpha^{(i)}_{\mathcal O} v^2}{2\Lambda^2}$, where $v$ corresponds to the vacuum expectation value: $v=250$ G$e$V. 

As has been remarked in \cite{deGouvea:2015euy, Deppisch:2015qwa}, the bounds for low sterile neutrino masses, coming from beam dump and rare LNV decays of mesons, are heavily dependent on the decay modes considered. As the effective lagrangian we are considering leads to the decay mode to a photon and an ordinary neutrino, those bounds do not apply in this work. For $m_N$ in the range above a few hundred G$e$V, the electroweak precision data involving lepton number violating processes put the most stringent bounds on the neutrino mixings, except for the coupling constants of the first fermions family where the most stringent limits come from $0\nu\beta\beta$-decay. For the second and third families, we find that the most restrictive are the $EWPD$ constraints. In the following we explain how these bounds are translated to the effective couplings $\alpha^i_{\mathcal{O}}$.

Some of the considered operators contribute directly to the neutrinoless double beta decay ($0\nu\beta\beta$-decay) and thus the corresponding coupling constants, involving the first fermion family $i=1$, are restricted by strong bounds. We explicitly calculated the implications for the effective couplings in our lagrangian.

In a general way, the following effective interaction Hamiltonian can be considered:
\begin{equation}
\label{eq:heff}
\mathcal{H}=G_{eff} \; \bar u \Gamma d \;\; \bar e
\Gamma N + h.c.
\end{equation}
where $\Gamma$ represents a general Lorentz-Dirac structure.
Following the development presented in \cite{Mohapatra:1998ye} we find 
\begin{equation*}
G_{eff} \leq A \times 10^{-8} \left( \frac{m_N}{100 GeV}
\right)^{1/2} GeV^{-2}
\end{equation*}
where the numerical constant $A$ depends on the nuclear model used and the lifetime for the $0\nu\beta\beta$-decay. We take the most 
stringent limit $\tau_{{0\nu}_{\beta\beta}} \geq 1.1 \times 10^{26}$ years obtained by the KamLAND-Zen Collaboration \cite{KamLAND-Zen:2016pfg}. 

In the effective theory we are considering the lowest order contribution to $0\nu_{\beta \beta}$-decay comes from the operators containing the $W$ field and the 4-fermion operators with quarks $u$, $d$, the lepton $e$ and the Majorana neutrino $N$. These operators contribute to the effective Hamiltonian \eqref{eq:heff}, with $G_{eff}=\frac{\alpha}{\Lambda^2}$ which, as we discussed in the paragraph after Eq.\eqref{lw_lz}, is related with the mixing angle between light and heavy neutrinos as  $U_{l_i N}^2=\left(\frac{\alpha v}{2 \Lambda^2}\right)^2$. We find that the value $A=3.2$ fits very well the bounds obtained for the mixings \cite{deGouvea:2015euy,Faessler:2014kka} in the literature. 

We can translate the limit coming from  $G_{eff}$ on $\alpha^{(1)}_{\mathcal{O}}$ which, for $\Lambda=1$ T$e$V, is
\begin{equation}\label{eq:0nubb}
\alpha^{bound}_{0\nu\beta\beta} \leq 3.2 \times 10^{-2}
\left(\frac{m_N}{100 ~GeV}\right)^{1/2}.
\end{equation}

On the other hand, to consider the existing bounds coming from collider, electroweak precision data ($EWPD$) and
Lepton-Flavor-Violating processes we define, following \cite{delAguila:2005pf}: $\Omega_{l l'} = U_{l N} U_{l' N}$
where the allowed values for the parameters are \cite{Bergmann:1998rg}:
\begin{eqnarray*}
\Omega_{e e}\leq 0.0054, \;\; \Omega_{\mu \mu}\leq 0.0096, \;\;
\Omega_{\tau \tau} \leq 0.016
\end{eqnarray*}

For the Lepton-Flavor-Violating process (LFV), e.g. $\mu \rightarrow e \gamma$, $\mu \rightarrow e e e$ and $\tau \rightarrow e e e$, which are induced by the quantum effects of the heavy neutrinos, we have several bounds \cite{Antusch:2014woa, Drewes:2015iva, Tommasini:1995ii, Alva:2014gxa} but the most restrictive one  comes from $Br(\mu\rightarrow e\gamma)\leq 5.7 \; 10^{-13}$\cite{Tommasini:1995ii, delAguila:2008pw}. This bound imposes $|\Omega_{e \mu}| \leq 0.0001$, and can be translated to the constants $\alpha$, being 
\begin{eqnarray*}
\Omega_{e \mu}=U_{e N} U_{\mu N} = \left(\frac{\alpha}2
\frac{v^2}{\Lambda^2} \right)^2 < 0.0001
\end{eqnarray*}
and for $\Lambda=1$ TeV we have
\begin{eqnarray}\label{eq:EWPD}
\alpha^{bound}_{EWPD} \leq 0.32
\end{eqnarray}

In order to simplify the discussion, for the numerical evaluation we only consider the two following situations. In the set we call {\bf A} the couplings associated to the operators that contribute to the $0\nu\beta\beta$-decay ($\mathcal{O}_{N e \phi}$, $\mathcal{O}_{d u N e}$, $\mathcal{O}_{Q u N L}$, $\mathcal{O}_{L N Q d}$ and  $\mathcal{O}_{Q N L d}$) for the first family are restricted to the corresponding bound $\alpha^{bound}_{0\nu\beta\beta}$ and the other constants are restricted to the bound determined by $EWPD$  $\alpha^{bound}_{EWPD}$. In the case of the set called {\bf B} all the couplings are restricted to the $0\nu\beta\beta$ bound $\alpha^{bound}_{0\nu\beta\beta}$ which is the most stringent. For the 1-loop generated operators we consider the coupling constant as $1/(16 \pi^2)$ times the corresponding tree-level coupling: $\alpha^{1-loop}=\alpha^{tree}/(16 \pi^2)$. Thus, for the operators $\mathcal{O}_{DN}$, $\mathcal{O}_{NW}$ and $\mathcal{O}_{\bar DN}$, which contribute to $0\nu\beta\beta$ we have 
\begin{equation*}
\alpha^{(1)}_{L_2}, \alpha^{(1)}_{L_3}, \alpha^{(1)}_{L_4} \sim \frac{1}{16\pi^2} \alpha^{bound}_{0\nu\beta\beta}
\end{equation*}
for fermions of the first family. For the remaining operators we take
\begin{equation*}
\alpha \sim \alpha^{bound}_{EWPD} , \alpha^{bound}_{0\nu\beta\beta} 
\end{equation*}
in the sets {\bf A} and {\bf B} respectively.

%


\section{FINAL REMARKS}\label{sec:fin} 

Searches for heavy neutrinos often rely on the possibility that they may decay to detectable particles. The interpretation of the corresponding results of such searches requires a model for the heavy neutrino decay. In this work we consider an effective approach for heavy Majorana neutrino interactions, and calculate the branching ratios for the different decay modes.

Depending on the Majorana neutrinos mass scale, the decay can have effects on different physical contexts like solar/astrophysical neutrinos, collider searches like the those taking place at the LHC, neutrino experiments as OPERA, MiniBoone, SHALON, etc. In particular, the effects on some of this experiments for low mass neutrinos were discussed in \cite{Duarte:2015iba}.

Summarizing,  we calculated the decay modes for the Majorana neutrinos $N$ in an effective theory approach. We presented the analytical results for the dominant channels,  discussed the existent bounds taken into account for the effective couplings, and displayed the different branching ratios and the total decay width for the heavy sterile neutrino considered.

{\bf Acknowledgments}

We thank CONICET (Argentina) and Universidad Nacional de Mar del
Plata (Argentina); and PEDECIBA, ANII, and CSIC-UdelaR (Uruguay) for their 
financial supports.

\appendix
\section{Complete effective lagrangian}

We present here the complete effective lagrangian obtained from the operators listed in \eqref{eq:ope1}, \eqref{eq:ope2} and \eqref{eq:ope3}. 
The full contributions are considered for the total Majorana neutrino decay width $\Gamma_{N}$. 

\begin{eqnarray}\label{leff_svb}
 \mathcal{L}^{tree}_{SVB} &= & \frac{1}{\Lambda^2}\left\{\alpha^{(i)}_{\phi}
\left( \frac{3v^2}{2\sqrt{2}}~\bar \nu_{L,i} N_R~ h + \frac{3v}{2\sqrt{2}}~\bar \nu_{L,i} N_R~ h h+ \frac{1}{2\sqrt{2}}~\bar \nu_{L,i} N_R~ h h h \right) \right. 
\nonumber
\\ && \left. - \alpha_Z \left( -(\bar N_R \gamma^{\mu} N_R) \left( \frac{m_Z}{v} Z_{\mu} \right) \left( \frac{v^2}{2} + v h + \frac{1}{2} h h \right) \right. \right.
\nonumber 
\\ &&  \left. \left. + (\bar N_R \gamma^{\mu} N_R) \left( \frac{v}{2} P^{(h)}_{\mu} h +\frac{1}{2} P^{(h)}_{\mu} h h \right)  \right) \right.
\nonumber
\\ && - \left. \alpha^{(i)}_W (\bar N_R \gamma^{\mu} l_R)\left(\frac{v m_{W}}{\sqrt{2}}W^{+}_{\mu} + \sqrt{2} m_{W} W^{+}_{\mu} h + \frac{g}{2 \sqrt{2}} W^{+}_{\mu} h h \right) + h.c. \right\}.
\end{eqnarray}

In \eqref{leff_svb} a sum over the family index $i$ is understood, and the constants $\alpha_{\mathcal{O}}^{(i)}$ are associated to the specific operators:
\begin{eqnarray*}
 \alpha_Z&=&\alpha_{NN\phi},\;
\alpha^{(i)}_{\phi}=\alpha^{(i)}_{LN\phi},\;
\alpha^{(i)}_W=\alpha^{(i)}_{Ne\phi}
\end{eqnarray*}.
The four-fermion contact operators presented in \eqref{eq:ope2} lead to the tree-level lagrangian:
\begin{eqnarray}\label{leff_4-f}
\mathcal{L}^{tree}_{4-f}&=& \frac{1}{\Lambda^2} \left\{ \alpha^{(i)}_{V_0} \bar d_{R,i} \gamma^{\mu} u_{R,i} \bar N_R \gamma_{\mu}
l_{R,i} + \alpha^{(i)}_{V_1} \bar l_{R,i} \gamma^{\mu} l_{R,i} \bar
N_R \gamma_{\mu} N_R + \alpha^{(i)}_{V_2} \bar L_i \gamma^{\mu} L_i
\bar N_R \gamma_{\mu} N_R + \right. \nonumber
\\ && \left. \alpha^{(i)}_{V_3} \bar u_{R,i} \gamma^{\mu}
u_{R,i} \bar N_R \gamma_{\mu} N_R + \alpha^{(i)}_{V_4} \bar d_{R,i}
\gamma^{\mu} d_{R,i} \bar N_R \gamma_{\mu} N_R + \alpha^{(i)}_{V_5}
\bar Q_i \gamma^{\mu} Q_i \bar N_R \gamma_{\mu} N_R + \right.
\nonumber
\\ && \left.
\alpha^{(i)}_{S_0}(\bar \nu_{L,i}N_R \bar e_{L,i}l_{R,i}-\bar
e_{L,i}N_R \bar \nu_{L,i}l_{R,i}) + \alpha^{(i)}_{S_1}(\bar
u_{L,i}u_{R,i}\bar N \nu_{L,i}+\bar d_{L,i}u_{R,i} \bar N e_{L,i})
 + \right. \nonumber
\\ && \left.
\alpha^{(i)}_{S_2} (\bar \nu_{L,i}N_R \bar d_{L,i}d_{R,i}-\bar
e_{L,i}N_R \bar u_{L,i}d_{R,i}) + \alpha^{(i)}_{S_3}(\bar u_{L,i}N_R
\bar e_{L,i}d_{R,i}-\bar d_{L,i}N_R \bar \nu_{L,i}d_{R,i}) + \right.
\nonumber
\\ && \left.  \alpha^{(i)}_{S_4} (\bar N_R \nu_{L,i}~\bar l_{L,i} N_R~+\bar
N_R e_{L,i} \bar e_{L,i} N_R)  + h.c. \right\}
\end{eqnarray}
In \eqref{leff_4-f} a sum over the family index $i$ is understood, and the
constants $\alpha^{(i)}_{\mathcal O}$ are associated to the specific operators:
\begin{eqnarray*}
\alpha^{(i)}_{V_0}&=&\alpha^{(i)}_{duNe},\;
\alpha^{(i)}_{V_1}=\alpha^{(i)}_{eNN},\;
\alpha^{(i)}_{V_2}=\alpha^{(i)}_{LNN},\; 
\alpha^{(i)}_{V_3}=\alpha^{(i)}_{uNN},\;
\alpha^{(i)}_{V_4}=\alpha^{(i)}_{dNN},\;
\alpha^{(i)}_{V_5}=\alpha^{(i)}_{QNN},\;
\;\nonumber \\
\alpha^{(i)}_{S_0}&=&\alpha^{(i)}_{LNe},\;
\alpha^{(i)}_{S_1}=\alpha^{(i)}_{QuNL},\;
\alpha^{(i)}_{S_2}=\alpha^{(i)}_{LNQd},\;\;
\alpha^{(i)}_{S_3}=\alpha^{(i)}_{QNLd},\;
\alpha^{(i)}_{S_4}=\alpha^{(i)}_{LN}.
\end{eqnarray*}

The one-loop level generated operators in \eqref{eq:ope3} give the lagrangian:
\begin{eqnarray}\label{leff_1loop}
\mathcal{L}_{eff}^{1-loop}&=&\frac{\alpha_{L_1}^{(i)}}{\Lambda^2} \left(-i\sqrt{2} v c_W P^{(A)}_{\mu} ~\bar \nu_{L,i} \sigma^{\mu\nu} N_R~ A_{\nu} 
+i \sqrt{2} v s_W P^{(Z)}_{\mu} ~\bar \nu_{L,i} \sigma^{\mu\nu} N_R~ Z_{\nu}+  \right.
\nonumber 
\\ && \left. -i\sqrt{2} c_W P^{(A)}_{\mu} ~\bar\nu_{L,i} \sigma^{\mu\nu} N_R~ A_{\nu} h + i \sqrt{2}  s_W P^{(Z)}_{\mu} ~\bar \nu_{L,i} \sigma^{\mu\nu} N_R~ Z_{\nu} h \right)  
\nonumber 
\\ && -\frac{\alpha_{L_2}^{(i)}}{\Lambda^2} \left(\frac{m_Z}{\sqrt{2}}P^{(N)}_{\mu} ~\bar \nu_{L,i} N_R~ Z^{\mu}+ \frac{m_{z}}{\sqrt{2}v} P^{(N)}_{\mu} ~\bar \nu_{L,i} N_R~ Z^{\mu} h  \right.
\nonumber
\\ &&  \left. + m_W P^{(N)}_{\mu} ~\bar l_{L,i} N_R~ W^{-\mu} + \frac{\sqrt{2} m_{W}}{v} P^{(N)}_{\mu} ~\bar l_{L,i} N_R~ W^{-\mu} h + \frac{1}{\sqrt{2}} P^{(h)}_{\mu} P^{(N)\mu} ~\bar \nu_{L,i} N_R~ h \right) 
\nonumber 
\\ && -\frac{\alpha_{L_3}^{(i)}}{\Lambda^2}\left(i\sqrt{2} v  c_W P^{(Z)}_{\mu} ~\bar \nu_{L,i} \sigma^{\mu\nu}N_R~ Z_{\nu} 
+ i\sqrt{2} v s_W P^{(A)}_{\mu} ~\bar \nu_{L,i} \sigma^{\mu\nu}N_R~ A_{\nu} \right.
\nonumber 
\\ && \left. +i 2\sqrt{2} m_W ~\bar \nu_{L,i} \sigma^{\mu\nu} N_R~ W^+_{\mu}W^-_{\nu} + i \sqrt{2} v P^{(W)}_{\mu} ~\bar l_{L,i} \sigma^{\mu\nu} N_R~ W^-_{\nu} \right.
\nonumber
\\ && \left. + i 4 m_W c_W ~\bar l_{L,i} \sigma^{\mu\nu} N_R~ W^-_{\mu} Z_{\nu}+ i 4 m_W s_W ~\bar l_{L,i} \sigma^{\mu\nu} N_R~ W^-_{\mu} A_{\nu} \right.
\nonumber 
\\ && \left. + i \sqrt{2} P^{(W)}_{\mu} ~\bar l_{L,i} \sigma^{\mu\nu} N_R~ W^-_{\nu} h + i 2 g  c_W ~\bar l_{L,i} \sigma^{\mu\nu} N_R~ W^-_{\nu} Z_{\mu} h \right.
\nonumber 
\\ && \left. + i 2 g s_W ~\bar l_{L,i} \sigma^{\mu\nu} N_R~ W^-_{\nu} A_{\mu} h  + i \sqrt{2} c_W P^{(Z)}_{\mu} ~\bar \nu_{L,i} \sigma^{\mu\nu} N_R~ Z_{\mu} h \right.
\nonumber
\\ && \left. + i \sqrt{2} s_W P^{(A)}_{\mu} ~\bar \nu_{L,i} \sigma^{\mu\nu} N_R~ A_{\mu} h +i \sqrt{2} g ~\bar \nu_{L,i} \sigma^{\mu\nu} N_R~ W^{+}_{\mu} W^{-}_{\nu} h
\right) 
\nonumber
\\ &&  -\frac{\alpha_{L_4}^{(i)}}{\Lambda^2} \left( \frac{m_Z}{\sqrt{2}} P^{(\bar\nu)}_{\mu}~\bar \nu_{L,i} N_R~ Z_{\mu}+ \frac{m_{Z}}{\sqrt{2}v} (P^{(\bar\nu)}_{\mu}-P^{(h)}_{\mu})~\bar \nu_{L,i} N_R~ Z^{\mu} h  \right. 
\nonumber
\\ && \left. + \frac{1}{\sqrt{2}} P^{(h) \mu} P^{(\bar\nu)}_{\mu}~\bar \nu_{L,i} N_R~ h -\frac{\sqrt{2} m^2_W}{v} ~\bar \nu_{L,i} N_R~  W^{-\mu}W^+_{\mu} 
-\frac{m^2_z}{\sqrt{2} v} ~\bar \nu_{L,i} N_R~ Z_{\mu}Z^{\mu}   \right.
\nonumber
\\ && \left. -\frac{1}{2}\frac{m^2_{Z}}{v^2} ~\bar \nu_{L,i} N_R~ Z_{\mu}Z^{\mu} h -\frac{\sqrt{2} m^2_W}{v^2} ~\bar \nu_{L,i} N_R~ W^{+}_{\mu} W^{-\mu} h \right.
\nonumber
\\ && \left. + m_W P^{(\bar l)}_{\mu} W^{-\mu} ~\bar l_{L,i} N_R~ +\frac{m_W}{v} (P^{(\bar l)}_{\mu}-P^{(h)}_{\mu}) W^{-\mu} ~\bar l_{L,i} N_R~ h \right.
\nonumber
\\ && \left. + e m_W   ~\bar l_{L,i} N_R~ W^{-\mu}A_{\mu}  + e m_Z s_W ~\bar l_{L,i} N_R~ W^{-\mu}Z_{\mu}  \right.
\nonumber
\\ &&  \left. + \frac{e m_Z s_W}{v} ~\bar l_{L,i} N_R~ Z_{\mu} W^{-\mu} h + \frac{e m_Z c_W}{\sqrt{2} v} ~\bar l_{L,i} N_R~ A_{\mu} W^{-\mu} h\right) + h.c.
\end{eqnarray}
where $P^{(a)}$ is the 4-moment of the incoming $a$-particle and a sum over the family index $i$ is understood again. 
The constants $\alpha^{(i)}_{L_j}$ with $j=1,2,3,4$ are associated to the specific operators:
\begin{eqnarray*}
\alpha^{(i)}_{L_1}=\alpha^{(i)}_{NB},\;\; \alpha^{(i)}_{L_2}=\alpha^{(i)}_{DN},\;\; \alpha^{(i)}_{L_3}=\alpha^{(i)}_{NW},\;\; 
\alpha^{(i)}_{L_4}=\alpha^{(i)}_{\bar DN}.
\end{eqnarray*} 

\section{Fermionic three body decay widths}\label{Fermionic}

The decays to one lepton and two quarks take the expressions:
\begin{eqnarray}\label{dquarks_1}
&&\frac{d\Gamma}{dx}^{(N\rightarrow l^{+} \bar u d)}=\frac{m_N }{512
\pi^3}\left(\frac{m_N}{\Lambda}\right)^4  x \frac{(1-x-y_l+y_u)}{(1-x+y_l)^ 3}
 \left\{ (1-x+y_l-y_u) \left[6 \alpha_1^{i_u,i_l} x (1-x+y_l)^2 \right. \right. \nonumber \\ &+& 
 12 \alpha_2^{i_u,i_l} (2-x)(1-x+y_l)\sqrt{y_l y_u}  +
\alpha_3^{i_u,i_l} (2 x^3-x^2(5+5 y_l+y_u)-4 y_l(1+y_l+2 y_u) \nonumber \\ &+&
\left.  x(3+10y_l+3y_l^2+3y_u+3y_l y_u)) \right] 
+ \left. 24 \alpha_4^{i_u,i_l} x (1-x+y_l)^2 \sqrt{y_l y_u} \right\}
\end{eqnarray} 
with $~2\sqrt{y_l}<x<1+y_l-y_u , ~ y_l={m_l^2}/{m_N^2}, ~y_u={m_u^2}/{m_N^2}$ 
and the coefficients $\alpha_{1,..,4}$ take the expressions:
\begin{eqnarray*}
\alpha_1^{i_u,i_l}&=&\left(\alpha_{s_1}^{(i_u)2}+\alpha_{s_2}^{(i_u)2}-
\alpha_{s_2}^{(i_u)}\alpha_{s_3}^{(i_u)} \right)\delta^{i_u,i_l}
\nonumber \\
\alpha_2^{i_u,i_l}&=&\left(\alpha_{s_1}^{(i_u)} \alpha_{W}^{(i_l)}\frac{y_W(1-x+y_l-y_W)}{(1-x+y_l-y_W)^2+y_W y_{\Gamma_W}}-
\alpha_{s_3}^{(i_u)}\alpha_{V_{0}}^{(i_u)}\right)\delta^{i_u,i_l}
\nonumber \\
\alpha_3^{i_u,i_l}&=&\left(\alpha_{s_3}^{(i_u)2}+4 \alpha_{V_{0}}^{(i_u)2}\right)\delta^{i_u, i_l}+
4 \alpha_{W}^{(i_l)2}\frac{y_W^2(1-x+y_l-y_W)}{(1-x+y_l-y_W)^2+y_W y_{\Gamma_W}}
\nonumber \\
\alpha_4^{i_u,i_l}&=& \alpha_{s_2}^{(i_u)} \alpha_{V_0}^{(i_u)} \delta^{i_u,i_l}
\end{eqnarray*}
\begin{eqnarray*}\label{dquarks_2}
\frac{d\Gamma}{dx}^{(N \rightarrow \nu d d)}&=&\frac{m_N}{128\pi^3}
\left( \frac{m_N}{\Lambda}\right)^4
\frac{x^2}{4}\frac{((1-x)(1-x-4y_d))^{1/2}}{(1-x)^2}
\left\{\delta^{i_l,i_d}\alpha_{s_3}^{(i_l)2} (3+x(-5+2 x+2 y_d))  \right. \nonumber \\  &+&  \left. 
6\left(\tilde{\alpha}_{1}^{i_l,i_d \, 2}+\tilde{\alpha}_{2}^{i_l,i_d \, 2}-\delta^{i_l,i_d}\tilde{\alpha}_{3}^{i_l,i_d}\alpha_{s_3}^{i_l}\right)
(1-x)(1-x-2 y_d) 
 \right\}
\end{eqnarray*}
with $0<x<1-4 y_d , ~y_d={m_d^2}/{m_N^2}$  and 
\begin{eqnarray*}
\tilde{\alpha}_{1}^{i_l,i_d \, 2}&&=\left(\delta^{i_l,i_d}\alpha_{s_2}^{(i_l)}+\alpha_{\phi}^{(i_l)} c \frac{(1-x-y_{h})}D \right)^2+
\alpha_{\phi}^{(i_l)2} c^2 \frac{y_{h} y_{\Gamma_{h}}}{D_{h}^2} 
\nonumber\\
\tilde{\alpha}_{2}^{i_l,i_d \, 2}&&=\alpha_{\phi}^{(i_l)} \frac{c^2}D_{h}
\nonumber\\
\tilde{\alpha}_{3}^{i_l,i_d}&&=\delta^{i_l,i_d}\alpha_{s_2}^{(i_l)} + \alpha_{\phi}^{(i_l)} ~c~ \frac{(1-x-y_{h})}{D_{h}}
\; , \; 
y_{h}={m_{h}^2}/{m_N^2} \; , \; 
y_{\Gamma_{h}}={\Gamma_{h}^2}/{m_N^2} 
\nonumber\\
D_{h}&&=(1-x-y_{h})^2+y_{h} y_{\Gamma_{h}} \; , \; c=\frac{3 g v^2 m_d}{4\sqrt{2} m_N^2 m_W}
\end{eqnarray*}.
\begin{eqnarray*}\label{dquarks_3}
\frac{d\Gamma}{dx}^{(N \rightarrow \nu u u)}&=&\frac{m_N}{128\pi^3}\left( \frac{m_N}{\Lambda}\right)^4  \tilde{\alpha}^{i_l,i_u \, 2}
\;\; \frac32 x^2 \left(1-\frac{4 y_u}{(1-x)}\right)^{1/2}(1-x-2 y_u) \delta_{i_u,i_l} 
\end{eqnarray*}
with $0<x<1-4 y_u$ and 
\begin{eqnarray*}
\tilde{\alpha}^{i_l,i_u \, 2}=\delta^{i_l,i_u} \alpha_{S_1}^{(i_l) \, 2}+ 2 \alpha_{\phi}^{(i_l) \, 2} \frac{c^2}{D_{h}} \; , 
\; c=\frac{3 g v^2 m_u}{4\sqrt{2} m_N^2 m_W}
\end{eqnarray*}.

The purely leptonic decay gives:
\begin{eqnarray*}\label{dleptons}
\frac{d\Gamma}{dx}^{(N\rightarrow l^{+} leptons)}&=&\frac{m_N }{1536\pi^3}
\left( \frac{m_N}{\Lambda} \right)^4 \frac{(1-x+y_l-y_{l^{\prime}})^2}{(1-x+y_l)^3} 
 x \left[ \alpha_1^{i_l,i_{l^{\prime}}} P(x)-\alpha_2^{i_l,i_{l^{\prime}}} R(x) \right] 
\end{eqnarray*}
with $2\sqrt{y_l} < x < 1+y_l-y_{l^{\prime}} ,~ y_l={m_l^2}/{m_N^2} , ~ y_{l^{\prime}}={m_l^{\prime 2}}/{m_N^2}$
and $\alpha_{1,2}$ and the terms $P(x)$, $R(x)$ take the expressions:
\begin{eqnarray*}
\alpha_1^{i_l,i_{l^{\prime}}}&=&\alpha_{s_0}^{(i_{l}) \, 2} \; \delta^{i_l,i_{l^{\prime}}}+
\frac{4 \alpha_W^{(i_l) \, 2} y_W^2}{(1-x+y_l-y_W)^2+y_W y_{\Gamma_W}}
\nonumber \\
\alpha_2^{i_l,i_{l^{\prime}}}&=&12 \alpha_{s_0}^{(i_{l^{\prime}})}\alpha_{W}^{(i_l)} 
\frac{(1-x+y_l-y_W)}{(1-x+y_l-y_W)^2+y_W y_{\Gamma_W}} \delta^{i_l,i_{l^{\prime}}}
\nonumber \\
P(x)&=&2x^3-x^2(5+5 y_l+y_{l^{\prime}})-4 y_l (1+y_l+2y_{l^{\prime}})+
x(3+10 y_l + 3 y_l^2+3y_{l^{\prime}}+3 y_l y_{l^{\prime}})
\nonumber \\
R(x) &=& (2-x)(1-x+y_l)(y_l y_{l^{\prime}})^{1/2}.
\end{eqnarray*}
%


\bibliography{Bib_N_06_2016}

\end{document}